\ifdefined\pdfoutput\pdfoutput=1\fi\documentclass{article}
\usepackage{iclr2026_conference,times}
\usepackage{amsmath,graphicx,hyperref}
\usepackage{adjustbox}
\usepackage{algorithm}
\usepackage{algorithmic}
\usepackage{multirow}
\usepackage{booktabs}
\usepackage{hhline}
\usepackage{array}
\usepackage{listings}
\usepackage[table]{xcolor}
\usepackage{amssymb}
\usepackage{threeparttable}
\usepackage{caption}
\usepackage{textcomp}\usepackage{tcolorbox}\newtcolorbox{promptbox}[1]{colback=blue!2!white,colframe=black!35,colbacktitle=blue!8!white,coltitle=black,fonttitle=\bfseries,title={#1},boxrule=0.4pt,arc=1mm,left=8pt,right=8pt,top=6pt,bottom=6pt,before skip=8pt,after skip=8pt,fontupper=\normalsize\raggedright}
\usepackage{placeins}
\usepackage{dsfont}
\usepackage{url}\usepackage{svg}
\usepackage{enumitem}\newcommand{\tableautodash}[1]{\noalign{\vskip0.22ex}\multispan{#1}\leaders\hbox{\rule{2pt}{0.35pt}\hskip2pt}\hfill\cr\noalign{\vskip0.22ex}}
\definecolor{TableHeader}{RGB}{240,242,247}
\definecolor{TableSection}{RGB}{210,231,241}
\definecolor{TableKeSpeech}{RGB}{210,231,241}
\definecolor{TableCVYue}{RGB}{216,239,196}
\definecolor{TableQwen}{RGB}{210,231,241}
\definecolor{TableStep}{RGB}{216,239,196}
\definecolor{TableSFT}{HTML}{F0F1FF}
\definecolor{TableRule}{HTML}{5F6570}
\definecolor{TableMuted}{HTML}{737A86}
\makeatletter
\newcommand{\tableheadvline}{\rule[\dimexpr-\dp\@arstrutbox+0.22ex\relax]{0.4pt}{\dimexpr\ht\@arstrutbox+\dp\@arstrutbox-0.44ex\relax}}
\makeatother
\newcommand{\tablesectionrule}{\specialrule{0.4pt}{0.22ex}{0.22ex}}
\newcommand{\tablegroupdash}[1]{%
  \noalign{\vskip0.22ex
    \hbox to\linewidth{\color{TableRule}%
      \leaders\hbox{\rule{2pt}{0.35pt}\hskip2pt}\hfill}%
    \vskip0.22ex}}

\title{Multimodal Conversational Context for LLM-Based ASR: Data Construction, Training, and Benchmark}

\author{\begin{tabular}[t]{@{}c@{}}%
Longhao Li, Jian Tang, Yuxiang Kong, Jie Chen, Binbin Zhang, Lei Xie\thanks{Corresponding author}, Xiangang Li\\[0.6ex]
\raisebox{-0.6ex}{\includegraphics[height=2.8ex]{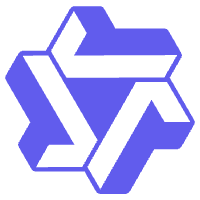}}\hspace{0.4em}\textbf{Alibaba Token Foundry}%
\end{tabular}}
\iclrfinalcopy

\begin{document}
\maketitle
\lhead{Technical Report}

\begin{abstract}
Conversational context provides semantic and acoustic cues across turns for automatic speech recognition (ASR), but relying on historical transcripts can propagate recognition errors and discard pronunciation and speaker information. We present a multimodal conversational-context framework for LLM-based ASR that integrates a scenario-controlled data pipeline, scalable multimodal context training, and systematic evaluation. We construct dialogues around entities and their confusable forms and interleave historical user speech with assistant text responses for supervised fine-tuning. We also introduce MM-ContextASR Bench, which evaluates contextual understanding and entity error correction across five scenarios. Experiments with Qwen3-Omni and Step-Audio-2-mini reveal limitations in handling irrelevant and erroneous history and show that our data construction and training improve context utilization, with multimodal context achieving the highest overall entity recall on both models. Further experiments on accent, dialect, and target-speaker ASR demonstrate the value of historical speech. The benchmark data and evaluation code are publicly available at \url{https://github.com/llh666521/MM-ContextASR}.
\end{abstract}

\section{Introduction}
Speech has become an important interface for interacting with intelligent assistants. Current spoken dialogue systems follow two main approaches: end-to-end systems process speech input directly to generate responses, whereas cascaded systems first convert speech into text using automatic speech recognition (ASR) and then use a text-based large language model (LLM) to generate responses. Cascaded architectures remain widely used in speech interaction systems, including recent systems such as DuplexCascade and DDTSR and the pipeline implementation in LiveKit Agents~\citep{yang2026duplexcascade,liu2026ddtsr,livekit2026agents}. Their modular design allows individual components to be optimized independently while reusing the reasoning and tool-use capabilities of text LLMs~\citep{liu2025xtalk}. In these systems, ASR connects user speech to the downstream language model. Recognition errors involving rare entities or domain-specific terms can alter the meaning of a request and propagate to subsequent understanding and response generation, making accurate recognition of such information essential.

Contextual ASR uses information beyond the current utterance to improve transcription. Traditional contextual biasing typically provides lists of candidate names, locations, or domain-specific terms to guide recognition~\citep{pundak2018deepcontext}. With the introduction of LLMs into ASR, recognizers can condition transcription on both speech and natural-language context, extending context use beyond predefined word lists~\citep{bai2024seedasr}. Prior work uses transcripts of earlier user utterances and textual assistant responses to provide entity, topic, and domain cues for recognizing the current turn~\citep{amazoncontextualasr}.

Textual dialogue history, however, is not always reliable or sufficient. Historical ASR transcripts may contain errors that assistant responses repeat, reinforcing incorrect entities in the context and interfering with current-turn recognition. Even when the transcript is correct, converting historical speech to text discards pronunciation, accent, and speaker characteristics. Multimodal context preserves the user's original speech, providing direct access to information that may be lost or corrupted in the transcript. This can reduce reliance on erroneous textual history and help mitigate error accumulation across turns. Historical acoustic features also complement textual semantics, providing cues for accent and dialect adaptation and target-speaker ASR.

% \begin{figure}[htbp]
%     \centering
%     \includesvg[width=0.65\linewidth,inkscapelatex=false]{figures/repeated_error_case.drawio}
%     \caption{A repeated-error case in which historical speech resolves an entity conflict that persists in the transcript and assistant response.}
%     \label{fig:repeated_error_case}
% \end{figure}

Motivated by these observations, we propose an LLM-based ASR framework with multimodal conversational context, encompassing a data construction pipeline, model training, and benchmark evaluation. The framework uses historical spoken user queries and textual assistant responses as context for current-turn recognition. Starting from an entity pool, the data pipeline performs confusion pair construction, dialogue generation, speech synthesis, and multimodal training data construction, producing conversations with varying contextual relevance and historical errors. Using these data, we interleave historical user speech and textual assistant responses in dialogue order, followed by the current user utterance, and apply supervised fine-tuning to generate the current-turn transcript from this multimodal context.

To systematically evaluate contextual understanding and error correction, we introduce MM-ContextASR Bench\footnote{\url{https://github.com/llh666521/MM-ContextASR}}. The benchmark holds the current speech and target entity fixed while varying the dialogue history across five controlled scenarios, using entity Recall to measure the effect of context on recognition. We further evaluate two tasks: accent and dialect adaptation on KeSpeech and CV-Yue~\citep{tang2021kespeech,ardila2020commonvoice}, and target-speaker ASR on AliMeeting~\citep{yu2022m2met}. These tasks assess the value of pronunciation and speaker cues in historical speech, respectively. Experimental results show that, after contextual fine-tuning, models benefit from historical speech and outperform text-only context in entity recall, accent and dialect recognition, and target-speaker ASR, demonstrating the advantages of multimodal conversational context.

\section{Related Work}
\noindent\textbf{LLM-based ASR.}
Recent LLM-based ASR systems incorporate the language modeling capabilities of LLMs into speech recognition, extending support for multilingual speech, dialects, and challenging acoustic conditions. Seed-ASR incorporates contextual information into recognition, while Qwen3-ASR supports multilingual and dialect recognition through large-scale speech training~\citep{bai2024seedasr,shi2026qwen3asr}. Qwen-Audio-3.0-ASR-Flash uses recently recognized content to improve contextual consistency across audio segments and supports specialized vocabulary recognition and on-the-fly hotword customization~\citep{qwen2026audio3asrflash}. Recent work also broadens the scope of recognition: MOSS-Transcribe-Diarize jointly produces transcripts, timestamps, and speaker labels, while AmphionASR explores personalized context-aware speech recognition~\citep{openmoss2026transcribe,amphion2026asr}. Industrial systems, including Doubao ASR and Hy ASR 3.0 preview, likewise address practical recognition needs such as dialects and specialized vocabulary~\citep{volcengine2026asr,tencent2026hyasr}.

\noindent\textbf{Conversational contextual ASR.}
Contextual ASR uses information beyond the current utterance to aid transcription, including predefined hotwords, domain prompts, and preceding dialogue~\citep{pundak2018deepcontext,bai2024seedasr}. Conversational contextual ASR focuses on the latter setting, using earlier turns to improve current-turn recognition. Prior studies explore dialogue history modeling to exploit topical continuity and lexical dependencies across turns~\citep{wei2023conversational}. However, recognition errors in historical transcripts can introduce misleading cues and propagate to subsequent predictions. To address this issue, prior work explores noise-aware training and preference optimization to improve robustness to unreliable context and mitigate mismatches between training- and inference-time histories~\citep{lee2024cnrl,guo2026noisycontext}. Our work focuses on multimodal user--assistant histories, using controlled data construction to vary contextual relevance and the presence of entity errors in historical transcripts and assistant responses, and to study how historical speech and text affect current-turn recognition.

\noindent\textbf{Contextual ASR Benchmarks.}
Recent benchmarks evaluate context utilization by controlling the information supplied to recognizers. ContextASR-Bench compares textual context at different levels of granularity, including domain and entity information~\citep{wang2025contextasrbench}. ProfASR-Bench evaluates professional speech using domain cues, speaker profiles, and adversarial prompts, with entity-aware measures of recognition performance~\citep{piskala2025profasr}. In contrast to these benchmarks centered on external textual prompts, MM-ContextASR Bench provides multimodal dialogue histories and controls their relevance and the occurrence of correct entities and confusable forms while holding the current utterance fixed. It uses entity Recall to evaluate contextual understanding and entity error correction.

\section{Multimodal Conversational Context ASR}
\subsection{Task Formulation}
We formulate ASR as current-turn transcription conditioned on multimodal dialogue history. Let $\mathbf{s}_t$, $\hat{\mathbf{y}}_t$, and $\mathbf{r}_t$ denote the user speech, its ASR hypothesis, and the assistant response at historical turn $t$, respectively. Given the current user speech $\mathbf{s}_T$, the model predicts only its reference transcript $\mathbf{y}_T$. We compare three matched context representations:
\begin{align}
\mathcal{C}_{\mathrm{text}} &= \{(\hat{\mathbf{y}}_t,\mathbf{r}_t)\}_{t=1}^{T-1}, \\
\mathcal{C}_{\mathrm{speech}} &= \{(\mathbf{s}_t,\mathbf{r}_t)\}_{t=1}^{T-1}, \\
\mathcal{C}_{\mathrm{speech+text}} &= \{(\mathbf{s}_t,\hat{\mathbf{y}}_t,\mathbf{r}_t)\}_{t=1}^{T-1}.
\end{align}
We refer to these representations as \emph{Text-only}, \emph{Speech-only}, and \emph{Speech+Text}. All three use the same current speech and reference transcript; only the historical representation changes. A fourth configuration, \emph{No Context}, removes the history and serves as the utterance-level ASR baseline.

\subsection{Scenario-Controlled Data Construction}
\label{sec:data_pipeline}
\begin{figure}[t]
    \centering
    \includesvg[width=\textwidth,inkscapelatex=false]{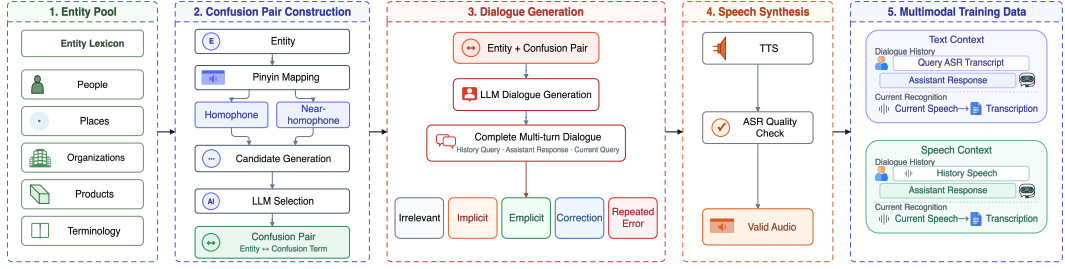}
    \caption{Our constructed-data pipeline. Starting from an entity lexicon, we construct acoustically plausible confusion pairs, generate complete multi-turn dialogues under five context scenarios, synthesize and verify speech, and package aligned text- and speech-context training examples.}
    \label{fig:data_pipeline}
\end{figure}

Figure~\ref{fig:data_pipeline} summarizes our five-stage pipeline for constructing user--assistant interactions with controlled relations between the history and current utterance.

\noindent\textbf{Entity Pool.}
We select about 175k target entities for training dialogue construction from an in-house lexicon designed for Mandarin ASR, with a focus on proper names and long-tail entities. These entities span domains such as medicine and healthcare, technology, finance, geography and tourism, and culture and entertainment, and include people, places, organizations, products, and technical terms.

\noindent\textbf{Confusion Pair Construction.}
We build a tone-independent Pinyin-to-character index from the entity lexicon. For each entity $e$, we retrieve homophonic replacement characters by matching Pinyin and near-homophonic alternatives through predefined initial or final substitutions, such as $n/l$, $s/sh$, and $in/ing$. Each candidate changes one character while preserving the rest of the entity. We retain up to 10 candidates per entity, prioritizing homophones. Qwen3.7-Max~\citep{qwen2026qwen37} then evaluates their lexical and semantic plausibility and selects a confusion term $\tilde{e}$ suitable as an ASR misrecognition, yielding the pair $(e,\tilde{e})$ for dialogue generation.

\noindent\textbf{Dialogue Generation.}
For each entity--confusion pair $(e,\tilde{e})$, we use Qwen3.7-Max to generate a current user query containing the correct entity $e$ and construct five types of historical context, each comprising one user query and an assistant response. The five scenarios share the same current query and control topical relevance and entity occurrences in the history to model topic shifts, topical continuity, and the handling of historical errors:
\begin{itemize}[leftmargin=*,nosep]
    \item \textbf{Irrelevant:} models a topic shift. The history and current query concern unrelated topics or domains, and the history contains neither $e$ nor $\tilde{e}$.
    \item \textbf{Implicit:} models continued discussion within a topic or domain. The history provides relevant semantic background without explicitly mentioning $e$ or $\tilde{e}$.
    \item \textbf{Explicit:} models successive questions about the same entity. The historical speech, simulated ASR transcript, and assistant response all contain the correct entity $e$, providing an explicit entity cue for the current query.
    \item \textbf{Correction:} models the assistant correcting a recognition error in its response. The historical speech contains $e$, the simulated ASR transcript replaces it with $\tilde{e}$, and the assistant response uses the correct entity $e$.
    \item \textbf{Repeated Error:} models a recognition error propagating into the assistant response. The historical speech contains $e$, but both the simulated ASR transcript and assistant response retain $\tilde{e}$, preserving the incorrect entity throughout the textual context.
\end{itemize}
In particular, for Correction and Repeated Error, we retain two textual versions of each historical query: one containing the correct entity $e$ for speech synthesis, and another containing $\tilde{e}$ as the simulated ASR transcript. For Irrelevant, we sample histories from previously generated dialogues about other entities and ensure that their topics are unrelated to the current query.

\noindent\textbf{Speech Synthesis.}
We synthesize the historical and current user queries as 24-kHz speech with Qwen-Audio~3.0 TTS~\citep{xiang2026audio30tts}; the synthesized audio is downsampled to 16 kHz for ASR. We then transcribe the synthesized speech using Paraformer~\citep{gao2022paraformer} and compare the resulting transcripts with the text submitted to TTS. To filter out low-quality synthesis, we use raw and text-normalized character error rates (CER), together with indicators of missing content, excessive insertions, and abnormal repetition.

\noindent\textbf{Multimodal Training Data.}
Each accepted conversation yields three aligned training views. Text-only contains the historical ASR hypothesis and assistant response; Speech-only replaces that hypothesis with the original historical speech; Speech+Text retains both representations together with the response. The current speech and supervision are identical across views, so the data isolate the contribution of the context modality.

\subsection{Multimodal Context Training}
\begin{figure}[t]
    \centering
    \includesvg[width=0.96\textwidth,inkscapelatex=false]{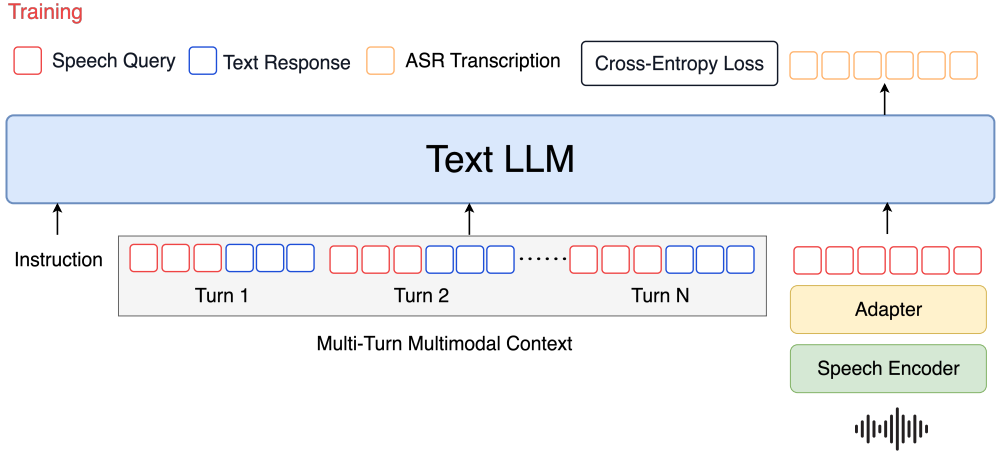}
    \caption{Our Speech-only multimodal context training scheme. Historical user speech queries and textual assistant responses are interleaved in dialogue order, followed by the current speech. Cross-entropy supervision is applied only to the current transcription.}
    \label{fig:training}
\end{figure}

Figure~\ref{fig:training} illustrates the Speech-only form of our context-training framework. The input begins with a task instruction and serializes the dialogue in its original order: historical user speech, the corresponding textual assistant response, and current user speech. The model then generates the transcript of the current utterance. Speech+Text inserts the historical ASR hypothesis after the corresponding speech segment, whereas Text-only replaces historical speech with that hypothesis.

The serialization supports multiple historical turns by repeating the user-speech and assistant-response blocks. We use the immediately preceding turn in this study because it provides the most local evidence and permits controlled comparison across tasks and context modalities. A shared speech encoder and modality adapter map historical and current speech into the LLM's embedding space. Let $\mathbf{c}$ denote the instruction and serialized history, and let $\mathbf{z}_T$ denote the encoded current speech. We minimize the autoregressive cross-entropy loss
\begin{equation}
    \mathcal{L}_{\mathrm{ASR}}
    = -\sum_{i=1}^{|\mathbf{y}_T|}
    \log p\!\left(y_{T,i}\mid \mathbf{c},\mathbf{z}_T,\mathbf{y}_{T,<i}\right).
\end{equation}
Loss positions corresponding to the instruction and dialogue history are masked; only tokens in the current-turn transcript contribute to $\mathcal{L}_{\mathrm{ASR}}$.

\FloatBarrier
\section{Evaluation Tasks}
We evaluate three forms of information carried by conversational history. MM-ContextASR Bench evaluates contextual understanding and entity error correction; KeSpeech and CV-Yue measure adaptation to Mandarin regional accents and Cantonese, respectively; and target-speaker ASR measures speaker localization in overlapping speech. For every task, we keep the current audio and reference transcript fixed and vary only the historical input among No Context, Text-only, Speech-only, and Speech+Text. This matched protocol attributes performance differences to the available context rather than to changes in the recognition target.

\subsection{MM-ContextASR Bench}
MM-ContextASR Bench evaluates contextual understanding and error correction through current-turn ASR across five controlled history scenarios. \textit{Irrelevant} tests whether the model can resist unrelated history, while \textit{Implicit} tests whether topic or domain cues improve transcription when the history contains neither the target entity nor its confusable form. \textit{Explicit} tests whether the model can associate a correctly mentioned historical entity with the current speech. \textit{Correction} and \textit{Repeated Error} introduce a confusable entity form into the historical transcript; the assistant response either corrects or retains it, respectively. These two scenarios assess whether the model can recover the correct entity rather than propagate historical errors, drawing on contextual cues, entity pronunciation in historical speech, and its linguistic knowledge.\par We construct it with the pipeline in Sec.~\ref{sec:data_pipeline}, using entity--confusion pairs held out before dialogue generation; neither the pairs nor the resulting conversations occur in the training data.

\begin{figure}[htbp]
    \centering
    \includegraphics[width=0.99\linewidth]{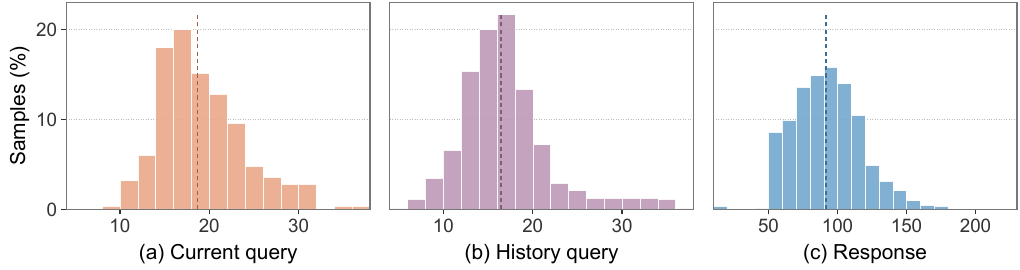}
    \caption{Text-length distributions in MM-ContextASR Bench: (a) current queries, (b) historical queries, and (c) assistant responses. Lengths are measured in characters after removing whitespace; dashed lines indicate means.}
    \label{fig:mm_contextasr_lengths}
\end{figure}

We manually select 250 target entities and use the data pipeline described above to construct dialogues across all five scenarios for each entity. Manual quality checks and validation yield 1,250 evaluation examples. The five scenarios for each entity share the same current speech and reference transcript, differing only in historical context. Each history comprises one user query and an assistant response, with the query provided as both speech and a simulated ASR transcript. Figure~\ref{fig:mm_contextasr_lengths} shows the text-length distributions of current queries, historical queries, and assistant responses.

We use target-entity recall as the primary metric, measuring the proportion of examples whose normalized current-turn transcription contains the canonical target entity. Because all five scenarios attached to an anchor share the same current utterance, No Context produces one prediction per anchor and therefore has identical recall across scenarios.

\subsection{Accent and Dialect Adaptation}
KeSpeech is a multi-accent Mandarin speech corpus containing about 1.5k hours of speech from 27k speakers, covering standard Mandarin and eight regional accents~\citep{tang2021kespeech}. We additionally evaluate Cantonese using the CV-Yue subset of Mozilla Common Voice~\citep{ardila2020commonvoice}. Because both corpora consist of isolated utterances rather than dialogue sessions, we create a pseudo-history by pairing each evaluation utterance with a different utterance from the same speaker. The same history--query pairs are used across context configurations. We evaluate the official KeSpeech test set of 19k utterances and the CV-Yue test set of 3.5k utterances with CER and sentence error rate (SER). CV-Yue transcripts are converted to simplified Chinese before training and scoring. We additionally report recall of LLM-annotated proper names and domain terms using exact matching, covering 3.7k mentions in 3.2k KeSpeech utterances and 395 mentions in 327 CV-Yue utterances.

\subsection{Target-Speaker ASR}
Target-speaker ASR requires the recognizer to transcribe a designated speaker while suppressing concurrent speech~\citep{zhang2023conformerTSASR,polok2025targetspeaker}. We evaluate this capability on AliMeeting, a Mandarin meeting corpus with synchronized far-field array recordings, close-talk recordings, and speaker-aware annotations~\citep{yu2022m2met}. From the official Eval partition, we extract 2.8k target segments whose far-field waveform contains speech from at least one competing speaker. For each segment, we select a temporally disjoint, non-overlapping utterance from the same speaker, meeting, and far-field channel as the reference history. The recognition reference contains only the target speaker's words, while the historical speech provides an acoustic reference for that speaker. The current waveform and target transcript remain identical across all four conditions.

\FloatBarrier
\section{Experiments} In our experiments, we seek to answer the following questions: \begin{itemize}[leftmargin=*,nosep] \item \textbf{(Q1) Contextual understanding and error correction.} How effectively do LLM-based ASR systems use textual and spoken dialogue history to recognize entities and correct errors, and how does contextual ASR fine-tuning improve this ability? \item \textbf{(Q2) Accent and dialect adaptation.} Can historical speech from the same speaker improve accent and dialect recognition, and does it provide benefits beyond textual context? \item \textbf{(Q3) Target-speaker ASR.} Can historical speech help LLM-based ASR systems identify and transcribe the target speaker in overlapping speech, and how does contextual ASR fine-tuning affect this capability? \end{itemize}
\subsection{Experimental Setup}

\begin{table}[t]
\centering
\caption{\textbf{Results on MM-ContextASR Bench.} Entity recall (\%) is reported for five anchor-aligned history scenarios. Base denotes the released model; Context SFT denotes condition-matched fine-tuning on our training set. Bold marks the best value in each metric column across Base and Context SFT for each model.}
\label{tab:bench_v1}
\setlength{\tabcolsep}{4.5pt}
\renewcommand{\arraystretch}{1.08}
\setlength{\aboverulesep}{0.22ex}
\setlength{\belowrulesep}{0.22ex}
\arrayrulecolor{black}
\normalsize\begin{adjustbox}{max width=\linewidth}
\begin{tabular}{@{}l|l|ccccc|c@{}}
\toprule
\textbf{Training} &
\textbf{Context} &
\textbf{Irrelevant $\uparrow$} & \textbf{Implicit $\uparrow$} & \textbf{Explicit $\uparrow$} & \textbf{Correction $\uparrow$} &
\textbf{Repeated Error $\uparrow$} &
\textbf{Overall $\uparrow$} \\
\tablesectionrule
\rowcolor{TableQwen}
\multicolumn{8}{c}{\rule{0pt}{2.1ex}\textbf{Qwen3-Omni-Instruct}} \\
\tablesectionrule
\multirow{4}{*}{Base}
 & No Context  & 74.00 & 74.00 & 74.00 & 74.00 & 74.00 & 74.00 \\
 & Text-only   & 68.40 & 74.00 & 95.60 & 96.00 & 67.60 & 80.32 \\
 & Speech-only & 65.60 & 74.40 & 92.40 & 96.40 & 72.00 & 80.16 \\
 & Speech+Text & 69.60 & 77.60 & 97.60 & 97.60 & 71.60 & 82.80 \\
\tableautodash{8}
\multirow{4}{*}{Context SFT}
 & No Context  & \textbf{76.80} & 76.80 & 76.80 & 76.80 & 76.80 & 76.80 \\
 & Text-only   & 76.40 & 83.20 & 97.20 & 98.40 & 77.20 & 86.48 \\
 & Speech-only & 76.40 & 82.40 & 98.40 & 98.40 & \textbf{82.80} & 87.68 \\
 & Speech+Text & 75.60 & \textbf{83.60} & \textbf{98.80} & \textbf{98.80} & 82.40 & \textbf{87.84} \\
\tablesectionrule
\rowcolor{TableStep}
\multicolumn{8}{c}{\rule{0pt}{2.1ex}\textbf{Step-Audio-2-mini}} \\
\tablesectionrule
\multirow{4}{*}{Base}
 & No Context  & 67.20 & 67.20 & 67.20 & 67.20 & 67.20 & 67.20 \\
 & Text-only   & 63.20 & 63.20 & 93.20 & 90.00 & 52.80 & 72.48 \\
 & Speech-only & 64.80 & 63.20 & 91.60 & 89.20 & 59.20 & 73.60 \\
 & Speech+Text & 64.40 & 66.40 & 95.60 & 91.20 & 53.20 & 74.16 \\
\tableautodash{8}
\multirow{4}{*}{Context SFT}
 & No Context  & \textbf{72.80} & 72.80 & 72.80 & 72.80 & 72.80 & 72.80 \\
 & Text-only   & 70.80 & 76.00 & 96.00 & 97.60 & 77.40 & 83.56 \\
 & Speech-only & \textbf{72.80} & 75.60 & \textbf{97.20} & \textbf{98.40} & 80.80 & 84.96 \\
 & Speech+Text & 72.00 & \textbf{76.80} & 96.80 & \textbf{98.40} & \textbf{82.00} & \textbf{85.20} \\
\bottomrule
\end{tabular}
\end{adjustbox}
\arrayrulecolor{black}
\end{table}
\noindent\textbf{Models and comparisons.}
We evaluate Qwen3-Omni-30B-A3B-Instruct~\citep{xu2025qwen3omni} and Step-Audio-2-mini~\citep{w2025stepaudio2} on MM-ContextASR Bench, and use Qwen3-Omni for all KeSpeech, CV-Yue, and AliMeeting experiments. \emph{Base} denotes the original model; \emph{Context SFT} denotes supervised fine-tuning on the corresponding task data. For each task and model, we train separate adapters for the four context configurations and evaluate each adapter under its matching input configuration.

\noindent\textbf{Training data.}
For MM-ContextASR, the pipeline in Sec.~\ref{sec:data_pipeline} produces about 873k training examples from 175k entity--confusion anchors across five history scenarios.\par For KeSpeech and CV-Yue, we fine-tune models separately on their respective official training splits, constructing same-speaker context pairs within each corpus as described in Sec.~4.2.\par For AliMeeting, we apply the target-speaker data construction described in Sec.~4.3 to the official Train partition, obtaining 80k training segments from about 200 meetings and 3k development segments from 10 disjoint meetings. We use the development set for periodic validation during training.

\noindent\textbf{Implementation details.}
We fine-tune the attention projections with LoRA~\citep{hu2022lora} (rank 64, scaling factor 128), keeping the speech encoder and modality adapter frozen. We use a learning rate of $1\times10^{-5}$, an effective batch size of 64, and a maximum sequence length of 8,192 tokens. Training runs for one epoch on MM-ContextASR and KeSpeech, and five epochs on CV-Yue and AliMeeting. All models are trained with ms-swift\footnote{\url{https://github.com/modelscope/ms-swift}} on eight NVIDIA A100 GPUs.

\noindent\textbf{Evaluation.} Appendix~\ref{app:prompts} provides the prompt templates and message layouts. We use deterministic decoding with fixed decoding parameters across context configurations within each task. All AliMeeting settings use the full 2.8k-example Eval set. We follow the task-specific metrics and scoring protocols in Sec.~4. Current audio, recognition references, and history selection remain fixed across context configurations.
\relax

\relax
\subsection{MM-ContextASR Bench}

\noindent\textbf{Contextual Limitations of Base Models.}
Table~\ref{tab:bench_v1} shows that both Base models underutilize conversational context. They benefit substantially from explicit correct-entity cues in Explicit, but gain little, or even lose ground, from topic or domain cues in Implicit, and consistently deteriorate with Irrelevant histories. The contrast between Correction and Repeated Error further reveals their dependence on correct entities in assistant responses: once the response repeats the historical error, Text-only falls below No Context. Even with historical speech containing the correct entity, neither model exceeds its No Context baseline in Repeated Error, and adding the erroneous transcript further reduces recall. Thus, accepting speech and text inputs does not by itself ensure effective understanding of multimodal dialogue history. MM-ContextASR Bench exposes the gap between exploiting explicit entity cues and interpreting context or recovering entities from erroneous histories, motivating targeted context training.

\noindent\textbf{Benefits of Multimodal Context.}
Preserving historical user speech allows conversational context to retain information beyond its transcript. After Context SFT, Speech+Text achieves the highest overall recall for both Qwen3-Omni and Step-Audio-2-mini, reaching 87.84\% and 85.20\%, compared with 86.48\% and 83.56\% for Text-only. The advantage is more pronounced in Repeated Error, with gains of 5.20 and 4.60 percentage points over Text-only, respectively. Here, both the historical transcript and assistant response repeat the incorrect entity, leaving no explicit correct-entity cue in the textual history. Historical speech instead preserves the original pronunciation and provides additional evidence for current-turn recognition. The value of multimodal conversational context therefore extends beyond higher average recall: it preserves information lost or misrepresented in transcription, supporting correct-entity recovery when historical text is unreliable.

\noindent\textbf{Effectiveness of Context SFT.}
Our context training enables more effective use of multimodal history for current-turn transcription. Before training, Speech+Text underperforms No Context in Repeated Error; after training, it exceeds the corresponding fine-tuned No Context baselines by 5.60 and 9.20 percentage points for Qwen3-Omni and Step-Audio-2-mini, respectively. This reversal shows that training translates previously underutilized historical speech into recognition gains despite erroneous textual context. The trained models also benefit more consistently from indirect cues in Implicit and incur smaller losses from unrelated histories in Irrelevant. Together with overall Speech+Text gains of 5.04 and 11.04 percentage points over Base, these results show that our training improves not only entity recognition accuracy but also the use of conversational history, extending context benefits beyond explicit correct-entity cues to indirect information and entity recovery under historical errors.

\FloatBarrier
\subsection{Accent and Dialect Recognition}

\begin{table}[!htbp]
\centering
\caption{\textbf{Accent and dialect ASR results.} We report recall, CER, and SER (\%) on KeSpeech and CV-Yue after simplified-Chinese normalization. Bold marks the best value in each metric column across Base and Context SFT for each dataset.}
\label{tab:accent_dialect_results}
% \small
\setlength{\tabcolsep}{4pt}
\renewcommand{\arraystretch}{1.08}
\setlength{\aboverulesep}{0.22ex}
\setlength{\belowrulesep}{0.22ex}
\arrayrulecolor{black}
\normalsize\begin{tabular}{@{}l|l|ccc@{}}
\toprule
\textbf{Training} &
\textbf{Context} &
\textbf{Recall $\uparrow$} & \textbf{CER $\downarrow$} & \textbf{SER $\downarrow$} \\
\tablesectionrule
\rowcolor{TableKeSpeech}
\multicolumn{5}{c}{\rule{0pt}{2.1ex}\textbf{KeSpeech}} \\
\tablesectionrule
\multirow{4}{*}{Base} & No Context  & 80.16 & 6.60 & 35.58 \\
& Text-only   & 78.78 & 6.88 & 36.54 \\
& Speech-only & 78.73 & 6.42 & 36.79 \\
& Speech+Text & 79.13 & 6.30 & 36.10 \\
\tableautodash{5}
\multirow{4}{*}{Context SFT} & No Context  & 83.71 & 4.51 & 29.56 \\
& Text-only   & 83.90 & 4.40 & 29.14 \\
& Speech-only & \textbf{84.93} & 4.19 & 28.39 \\
& Speech+Text & \textbf{84.93} & \textbf{4.17} & \textbf{28.20} \\
\tablesectionrule
\rowcolor{TableCVYue}
\multicolumn{5}{c}{\rule{0pt}{2.1ex}\textbf{CV-Yue}} \\
\tablesectionrule
\multirow{4}{*}{Base} & No Context  & 85.32 & 4.59 & 30.72 \\
& Text-only   & 81.01 & 5.10 & 33.67 \\
& Speech-only & 81.27 & 4.95 & 32.68 \\
& Speech+Text & 83.04 & 4.79 & 32.28 \\
\tableautodash{5}
\multirow{4}{*}{Context SFT} & No Context  & 84.81 & 4.89 & 33.08 \\
& Text-only   & 85.06 & 4.14 & 28.51 \\
& Speech-only & \textbf{87.85} & 3.83 & 26.36 \\
& Speech+Text & \textbf{87.85} & \textbf{3.74} & \textbf{26.33} \\
\bottomrule
\end{tabular}
\arrayrulecolor{black}
\end{table}

Table~\ref{tab:accent_dialect_results} shows that conversational history provides both linguistic and acoustic references for accent and dialect recognition. Text-only preserves the same speaker's previous word choices and expressions, which can inform current-turn transcription. Speech-only and Speech+Text additionally retain pronunciation, accent, and other speaking characteristics that the transcript does not directly convey. After Context SFT, Speech+Text achieves the lowest CER on both KeSpeech and CV-Yue, at 4.17\% and 3.74\%, compared with 4.40\% and 4.14\% for Text-only. Speech-only and Speech+Text both achieve entity recall of 84.93\% on KeSpeech and 87.85\% on CV-Yue, exceeding Text-only by 1.03 and 2.79 percentage points, respectively. The advantage of speech-conditioned input supports the value of a pronunciation reference beyond textual history. Context SFT makes these cues more useful: on CV-Yue, all contextual configurations underperform No Context before training, whereas all improve upon the corresponding No Context model after training. Thus, even when historical transcripts are accurate, retaining the original speech provides additional information for accent and dialect adaptation.

\FloatBarrier
\subsection{Target-Speaker ASR}

\begin{table}[!htbp]
\centering
\caption{\textbf{Target-speaker ASR results.} CER and SER (\%) are measured on 2.8k overlapping far-field segments from AliMeeting Eval. Bold marks the best value in each metric column across Base and Context SFT for each dataset.}
\label{tab:alimeeting_target_speaker}
% \small
\setlength{\tabcolsep}{4pt}
\renewcommand{\arraystretch}{1.08}
\setlength{\aboverulesep}{0.22ex}
\setlength{\belowrulesep}{0.22ex}
\arrayrulecolor{black}
\normalsize\begin{tabular}{@{}l|l|cc@{}}
\toprule
\textbf{Training} &
\textbf{Context} &
\textbf{CER $\downarrow$} & \textbf{SER $\downarrow$} \\
\tablesectionrule
\rowcolor{TableSection}
\multicolumn{4}{c}{\rule{0pt}{2.1ex}\textbf{AliMeeting}} \\
\tablesectionrule
\multirow{4}{*}{Base} & No Context  & 29.70 & 82.74 \\
& Text-only   & 31.87 & 82.98 \\
& Speech-only & 33.26 & 81.30 \\
& Speech+Text & 32.23 & 81.40 \\
\tableautodash{4}
\multirow{4}{*}{Context SFT} & No Context  & 30.25 & 75.61 \\
& Text-only   & 29.60 & 74.60 \\
& Speech-only & \textbf{24.59} & 72.18 \\
& Speech+Text & 24.80 & \textbf{71.96} \\
\bottomrule
\end{tabular}
\arrayrulecolor{black}
\end{table}

Table~\ref{tab:alimeeting_target_speaker} shows the role of historical speech as an identity reference in overlapping audio. Base obtains 31.87\%, 33.26\%, and 32.23\% CER for Text-only, Speech-only, and Speech+Text, respectively, all above No Context (29.70\%). Speech-conditioned Base inputs nevertheless reduce SER from 82.74\% to 81.30\%/81.40\%, indicating that their benefit depends on the metric rather than a complete failure to use reference speech. After Context SFT, Speech-only and Speech+Text achieve 24.59\% and 24.80\% CER, compared with 30.25\% for No Context and 29.60\% for Text-only. Historical text supplies linguistic context but not a direct voice reference; the stronger speech-conditioned results support learning to select the target voice in the current mixture. Speech-only achieves the lowest CER, while Speech+Text achieves the lowest SER (71.96\%). Thus, historical speech supports speaker selection in addition to the linguistic and pronunciation cues examined in the other tasks.

\FloatBarrier
\section{Conclusion}
We presented a unified framework for multimodal conversational context in LLM-based ASR, encompassing a scenario-controlled data pipeline, a training procedure applicable across context modalities and ASR tasks, and systematic evaluation. MM-ContextASR Bench reveals that Base models benefit from explicit entity cues but struggle to exploit indirect context, resist historical errors, and use historical speech effectively. Our proposed data construction and multimodal context training improve the models' ability to use conversational history, with Speech+Text achieving the highest overall entity recall on both models. Experiments on KeSpeech and CV-Yue further demonstrate the value of historical speech for accent and dialect adaptation, while AliMeeting results show its role as a reference for target-speaker recognition. Together, these findings establish the value of retaining historical user speech alongside textual dialogue context and training models to use its semantic, pronunciation, and speaker information for current-turn transcription.

\bibliographystyle{iclr2026_conference}
\bibliography{strings,refs}

\clearpage\appendix
\section{Prompt Templates}
\label{app:prompts}
English translations of the Chinese prompts used in our experiments are shown below. Audio placeholders distinguish historical and current speech; brace-delimited fields are filled for each example. All settings share the following system message.

\begin{promptbox}{Shared system message}
You are a speech recognition assistant.
\end{promptbox}

\subsection{MM-ContextASR Bench}

\begin{promptbox}{No Context}
\textbf{User:}\newline
Current audio: \textless{}current\_audio\textgreater{}\newline
Transcribe the current audio. Output only the transcription.
\end{promptbox}

\begin{promptbox}{Text-only}
\textbf{User:}\newline
\{history\_query\_transcript\}
\par\smallskip
\textbf{Assistant:}\newline
\{history\_response\}
\par\smallskip
\textbf{User:}\newline
Current audio: \textless{}current\_audio\textgreater{}\newline
Transcribe the current audio using the dialogue history as context. Output only the transcription.
\end{promptbox}

\begin{promptbox}{Speech-only}
\textbf{User:}\newline
\textless{}history\_audio\textgreater{}
\par\smallskip
\textbf{Assistant:}\newline
\{history\_response\}
\par\smallskip
\textbf{User:}\newline
Current audio: \textless{}current\_audio\textgreater{}\newline
Transcribe the current audio using the dialogue history as context. Output only the transcription.
\end{promptbox}

\begin{promptbox}{Speech+Text}
\textbf{User:}\newline
\textless{}history\_audio\textgreater{}\newline
\{history\_query\_transcript\}
\par\smallskip
\textbf{Assistant:}\newline
\{history\_response\}
\par\smallskip
\textbf{User:}\newline
Current audio: \textless{}current\_audio\textgreater{}\newline
Transcribe the current audio using the dialogue history as context. Output only the transcription.
\end{promptbox}

\subsection{KeSpeech and CV-Yue}
\begin{promptbox}{No Context}
\textbf{User:}\newline
Current audio: \textless{}current\_audio\textgreater{}\newline
Transcribe the current audio. Output only the transcription.
\end{promptbox}
\begin{promptbox}{Text-only}
\textbf{User:}\newline
The reference transcript below and the current audio are from the same speaker. Use the vocabulary, topics, and expressions in the reference transcript to help transcribe the current audio.\newline
Reference transcript: \{history\_text\}\newline
Current audio: \textless{}current\_audio\textgreater{}\newline
Transcribe the current audio. Output only the transcription.
\end{promptbox}
\begin{promptbox}{Speech-only}
\textbf{User:}\newline
The reference audio and the current audio are from the same speaker. Use the pronunciation patterns and accent in the reference audio to help transcribe the current audio.\newline
Reference audio: \textless{}history\_audio\textgreater{}\newline
Current audio: \textless{}current\_audio\textgreater{}\newline
Transcribe the current audio. Output only the transcription.
\end{promptbox}
\begin{promptbox}{Speech+Text}
\textbf{User:}\newline
The reference audio and the current audio are from the same speaker. Use the pronunciation patterns and accent in the reference audio to help transcribe the current audio.\newline
Reference audio: \textless{}history\_audio\textgreater{}\newline
Reference transcript: \{history\_text\}\newline
Current audio: \textless{}current\_audio\textgreater{}\newline
Transcribe the current audio. Output only the transcription.
\end{promptbox}

\subsection{AliMeeting}
\begin{promptbox}{No Context}
\textbf{User:}\newline
Speech transcription: \textless{}current\_audio\textgreater{}
\end{promptbox}
\begin{promptbox}{Text-only}
\textbf{User:}\newline
Reference transcript: \{reference\_text\}\newline
Use this text as a reference to help transcribe the target speaker in the current multi-speaker audio.\newline
Transcribe only this speaker's speech.\newline
Speech transcription: \textless{}current\_audio\textgreater{}
\end{promptbox}
\begin{promptbox}{Speech-only}
\textbf{User:}\newline
Use the speaker characteristics in the reference audio to identify the same speaker in the current multi-speaker audio, and transcribe only that speaker's speech.\newline
Reference audio: \textless{}history\_audio\textgreater{}\newline
Speech transcription: \textless{}current\_audio\textgreater{}
\end{promptbox}
\begin{promptbox}{Speech+Text}
\textbf{User:}\newline
Use the speaker characteristics in the reference audio to identify the same speaker in the current multi-speaker audio, and transcribe only that speaker's speech.\newline
Reference audio: \textless{}history\_audio\textgreater{}\newline
Reference transcript: \{reference\_text\}\newline
Speech transcription: \textless{}current\_audio\textgreater{}
\end{promptbox}
\end{document}